# *In Silico* Approaches to Deliver Better Antibodies by Design – The Past, the Present and the Future

Andreas Evers*, Shipra Malhotra*, Vanita D. Sood*

Andreas.Evers@merckgroup.com
Shipra.Malhotra@takeda.com
vanitadevisood@gmail.com

**Abstract**

The recognition of the importance of drug-like properties beyond potency to reduce clinical attrition of biologics has driven significant progress in the development of *in vitro* and *in silico* tools for developability assessment of antibody sequences. It is now routine to identify and eliminate or optimize antibody hits with poor developability profiles. To further accelerate discovery timelines and reduce clinical and non-clinical development attrition rates, more proactive *in silico* approaches to design sequence spaces with favorable developability profiles are required. From pragmatically front-loading structure based drug design for developability, to combining next generation sequencing with machine learning to shape screening libraries, to adapting the use of artificial intelligence and deep learning for immunoglobulins, we review herein progressively more proactive approaches to developability by design.

**Abbreviations**:
Anti-drug Antibodies (ADA)
Artificial Intelligence (AI)
Chemistry Manufacturing & Control (CMC)
Critical Quality Attribute (CQA)
Complementarity Determining Region (CDR)
Deep Learning (DL)
Denoising autoencoder (DAE)
Drug Metabolism and Pharmacokinetics (DMPK)
Fluorescence activated cell sorting (FACS)
Generative Adversarial Network (GAN)
Low-density lipoprotein (LDL)
Machine Learning (ML)
Major Histocompatibility Complex class II (MHC-II)
Multiple Sequence Alignment (MSA)
New Chemical Entity (NCE)
Next-generation sequencing (NGS)
Post-translational modification (PTM)
Protein Data Bank (PDB)
Sequence-activity-relationship (SAR)
Variational Auto-Encoder (VAE)

**Box 1. Defining the entities from hit to development candidate**

**Developability**: In a broad sense, developability is defined as the likelihood that an antibody sequence can become a manufacturable, safe and efficacious drug. To be considered developable, a lead should have a potency, exposure and formulatability compatible with a reasonable dose; low immunogenicity and low toxicity at the recommended dose; and biophysical properties that ensure a chemistry, manufacturing and control (CMC) process at reasonable costs within a reasonable timeline.

**Hit**: A sequence discovered from a screening campaign that displays the desired biochemical activity, target specificity, and mechanism of action (MoA). Typically requires extensive optimization in order to be considered developable.

**Optimized Hit**: A sequence optimized hit, that meets all the biochemical and cellular screening criteria and in addition is more drug-like in sequence, i.e. displays fewer liabilities and potential liabilities.

**Lead**: A sequence optimized hit that has additionally shown *in vivo* pharmacological activity in one or more disease models, and demonstrates acceptable pharmacokinetics (PK), non-clinical safety and manufacturability in preliminary assessments. Typically, a small pool of lead molecules will exist from which a development candidate and a backup candidate will be nominated.

**Development Candidate:** A stable transfection clone of a lead molecule which has undergone scale-up, rigorous *in vivo* testing, formal manufacturability assessment and passed all preclinical pharmacological, PK, non-clinical safety and CMC criteria to move into early non-clinical development (e.g. pivotal toxicity study) and eventually into clinical development.

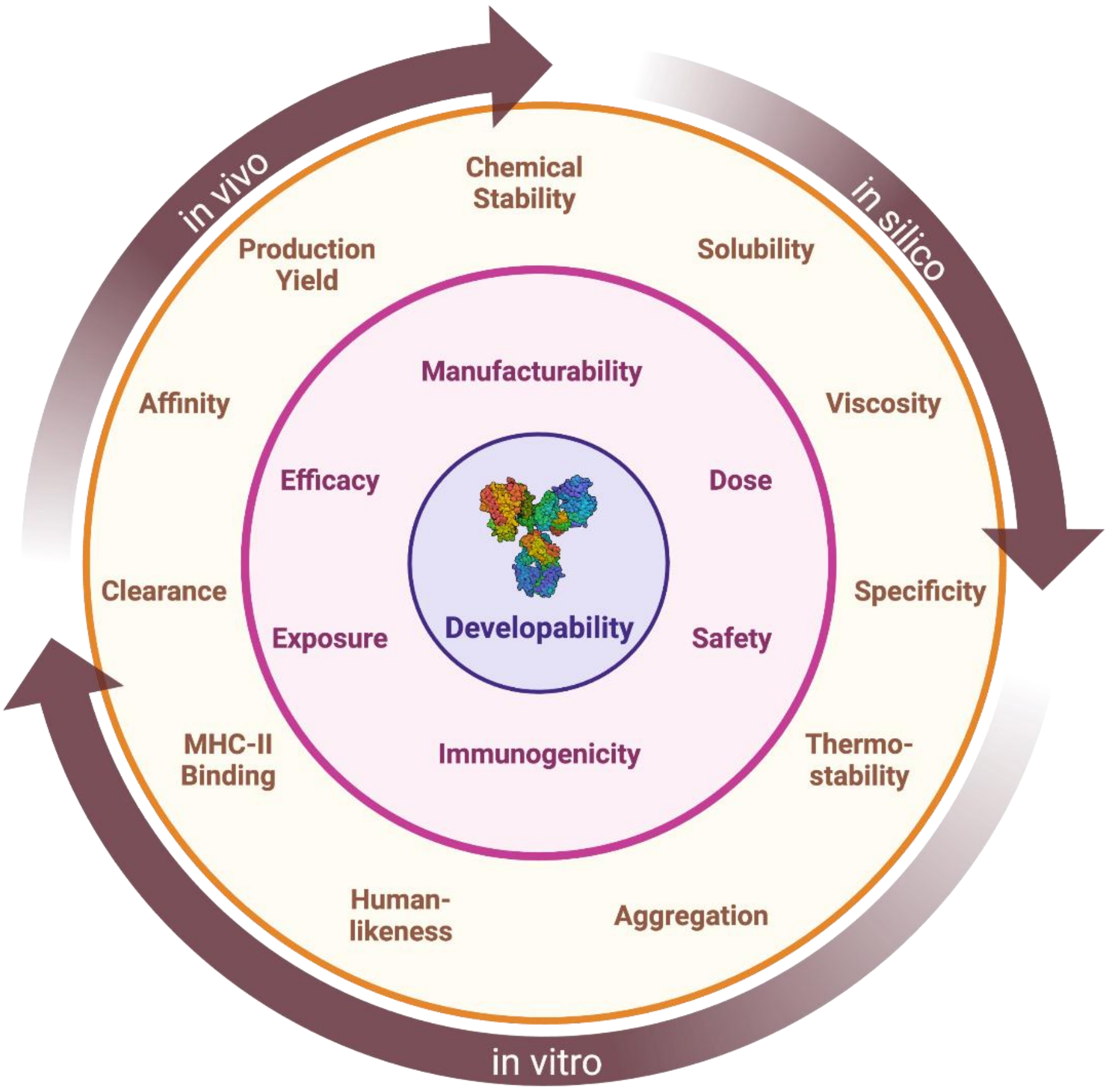

## 9.1 Introduction

Much ink has been spilled bemoaning the costs, both to drug developers and to society, of poorly developable drugs [1–4], and rightfully so as every delay and termination of what could have otherwise been an efficacious treatment for patients in need is consequential. Although biologics enjoy a somewhat higher rate of approval than new chemical entities [3,5–7], there remains substantial room for improvement. Invisible to the general public is the lost potential of many drug candidates that never enter clinic, due to decisions by drug developers to shelve compounds with poor developability prior to clinical trials, often after most or all of the preclinical discovery and early non-clinical development costs have already been incurred. Yet the consequences of moving forward with poorly developable drugs are even more harmful, with potential fallout ranging from delays in progression of clinical trials due to difficulties in producing enough drug supply, or to adverse events resulting from the intrinsic properties of the drug itself, its degradation products, or impurities associated to inconsistencies in manufacturing drugs with poor developability [8–13]. While decisions to terminate trials are typically made for multiple scientific and business reasons that are difficult to deconvolute (for an excellent review, see [14], there are some illustrative examples where molecular developability characteristics played a major role.

Bococizumab was a promising PCSK9 inhibitory monoclonal antibody with low-density lipoprotein (LDL) lowering activity that was investigated in cardiovascular indications as a possible statin alternative. The clinical development of bococizumab was stopped by the sponsor after considerable investment in eight phase 3 studies [15]. Anti-drug antibodies (ADA) were observed in only 7% of patients during phase 2 [16], whereas the larger phase 3 trial revealed 50% rate of ADA, with 29% of patients developing neutralizing ADA [17] leading to reduced efficacy [12]. While commercial factors certainly played a role in the decision to discontinue (alirocumab and evolocumab, in-class competitors, had already achieved market authorization), as well as variable efficacy even among patients without ADA [12], developability factors were also cited, specifically the high rate of ADA formation, as a major factor in the termination. ATR-107, an anti-IL21 monoclonal provides another example of a clinical trial halted, at least in part, due to a high rate of ADA formation [10].

CMC influencing biophysical properties have been a frequent target of post-hoc computational model informed engineering in the literature. Examples include CNTO607, an anti-IL13 monoclonal [9,13]; an anti-CD3ε antibody [18]; and an anti-VEGF antibody [8] for which a dramatic improvement of biophysical properties could be demonstrated. In the case of anti-VEGF, not only were solution properties improved, expression titer and product quality also improved substantially. Stamulumab (MYO-029) provides an additional compelling example; this monoclonal underwent a Phase I/II trial in muscular dystrophy patients that demonstrated a good safety profile, but no statistically significant effect was observed in exploratory efficacy endpoints [19], likely due to insufficient exposure for efficacy [20]. In addition to intrinsic clearance properties, both dose and route of administration can significantly affect exposure. In retrospect, it seems likely that the low solubility, high viscosity and aggregation propensity of stamulumab [21] would preclude a formulation and dose that would be needed to support efficacy, given the high free concentration of drug required to observe pharmacodynamic effects [20]. Initial post-hoc developability engineering demonstrated encouraging trends towards improved solubility [21], and subsequent model guided optimization yielded remarkable improvements in biophysical properties accompanied by potency increases [11]. Had the optimized variant been available, it is conceivable that higher doses could have been considered for clinical trials.

These examples of clinical stage developability failures explain the pharmaceutical industry's intense focus on early identification of drug candidates with improved developability characteristics (see Box 1 for definitions of developability and candidates, as used in this chapter). As full developability analyses cannot be performed until late program stages, early mitigation is a major challenge; yet this challenge has been met and overcome in small molecule discovery, providing a heartening example to follow; in the 1990s, most clinical attrition of small molecule drugs was due to poor developability, primarily poor exposure [3]; today, very few small molecules fail due to insufficient exposure in clinic, as modern discovery processes deliver developable candidates with good pharmacokinetics. Improved developability of biologics can in principle be achieved in one of two ways. The first option is to identify and remove from further development any screening hits that are more likely to display developability issues later in development. This option requires the ability to discriminate candidates that have liabilities. Many experimental [22–28] and computational [29–39] methods of candidate assessment are available and have been previously discussed at length [40–43]. The second option is to generate drug candidates de-risked for developability by design. This option requires early hit discovery and optimization methods that prioritize developability properties equally with on-target potency and functional efficacy. In both cases biopharmaceutical informatics and *in silico* engineering can have a significant impact.

Both options require an understanding of which preclinically measurable properties of the molecule actually predict the clinical developability criteria; the predictivity is not the same for each criterion and for each property. For example, it is well accepted that human and humanized sequences have lower immunogenicity on average than the murine and chimeric antibodies of old [44]. Predicting clinical immunogenicity for a specific drug candidate, however, remains challenging, as it is a complex phenomenon that springs from aspects of sequence (MHC-II presentation and proteosomal processing), and manufacturing process (levels of impurities, high or low molecular weight species, host cell protein contamination). There is no single *in silico* or *in vitro* assay that is holistically predictive of immunogenicity. The best we can do is to de-risk as much as possible, based on the predictive models that are available to us. The situation is similar for predicting expression yields, clearance and pharmacokinetics. On the other hand, developability aspects concerned with manufacturing criteria such as viscosity and route of administration are becoming much more predictable preclinically. Furthermore, a growing body of evidence suggests that even for developability properties that are more difficult to predict accurately, optimization on properties that are more easily measured *in silico* and *in vitro*, such as hydrophobicity, polyreactivity or pI, are likely to positively affect developability [23,25,29,45–48].

In this chapter we will focus on proactive and early computational de-risking approaches. We have divided the approaches into three broad categories we have designated as “classical”, “contemporary”, and “emerging” (Figure 1). The classical approach entails applying readily available and standard tools of sequence and structure-based drug design to rapidly & pragmatically re-engineer antibody hits obtained from traditional hit discovery approaches. The contemporary approach is more proactive in that it entails using next generation sequencing (NGS) and machine learning (ML) to engineer the libraries or repertoires from which hits are obtained for improved developability, thus aiming to ensure that every drug candidate selected on the basis of potency will by default have minimal developability liabilities. Finally, the emerging approach goes one step further, taking advantage of recent advances in deep learning (DL) and artificial intelligence (AI) in the field of protein structure prediction, and utilizes *de novo* computational design to directly generate drug candidates that meet multiple criteria for potency and developability.

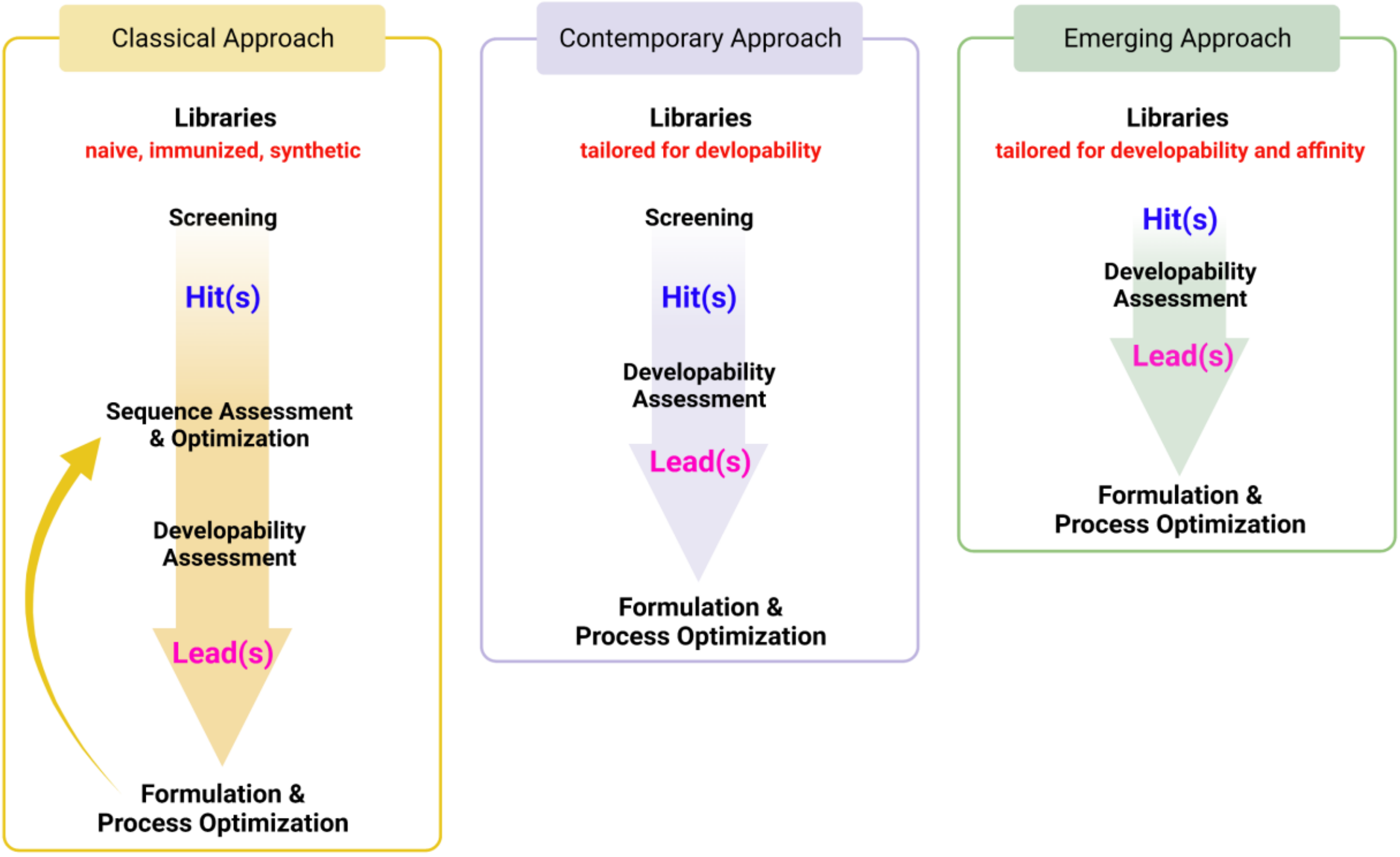

Figure 1. Illustration of the classical, contemporary and emerging approaches. The classical approach starts from a pool of potent hits from screening naïve, immunized or synthetic libraries that typically require cycles of sequence optimization. The contemporary approach utilizes engineered libraries that have been tailored by design towards improved developability properties and will ideally produce hits from the experimental screening campaign that will need minimal to no further sequence optimization. Finally, the emerging approach utilizes AI/ML approaches for a *de novo in silico* design of sequences optimized for developability and manufacturability.

### 9.2 The classical approach - Design of specific sequence-optimization variants

A traditional screening cascade for antibody discovery involves screening a repertoire (either synthetic or derived from a naïve or immunized subject) for a few key properties related to the desired on-target effect, e.g., binding to the target and not the off-target(s), ligand competition, and functional or phenotypic assay (Figure 2). After a handful of hits (sometimes as few as 2 or 3) that best meet the criteria are identified, these are prioritized for hit to lead optimization, typically including affinity maturation (if needed), humanization (if not derived from a human repertoire) and optimization of CDRs to remove chemically labile residues. The optimized hits are then scaled-up and tested *in vivo* for efficacy and clearance, and in early tests of manufacturability to select a lead and backup. With luck, at least one lead and/or backup will continue to demonstrate efficacy, tolerability and manufacturability as they are tested in increasingly rigorous preclinical studies, and will meet all criteria for a development candidate.

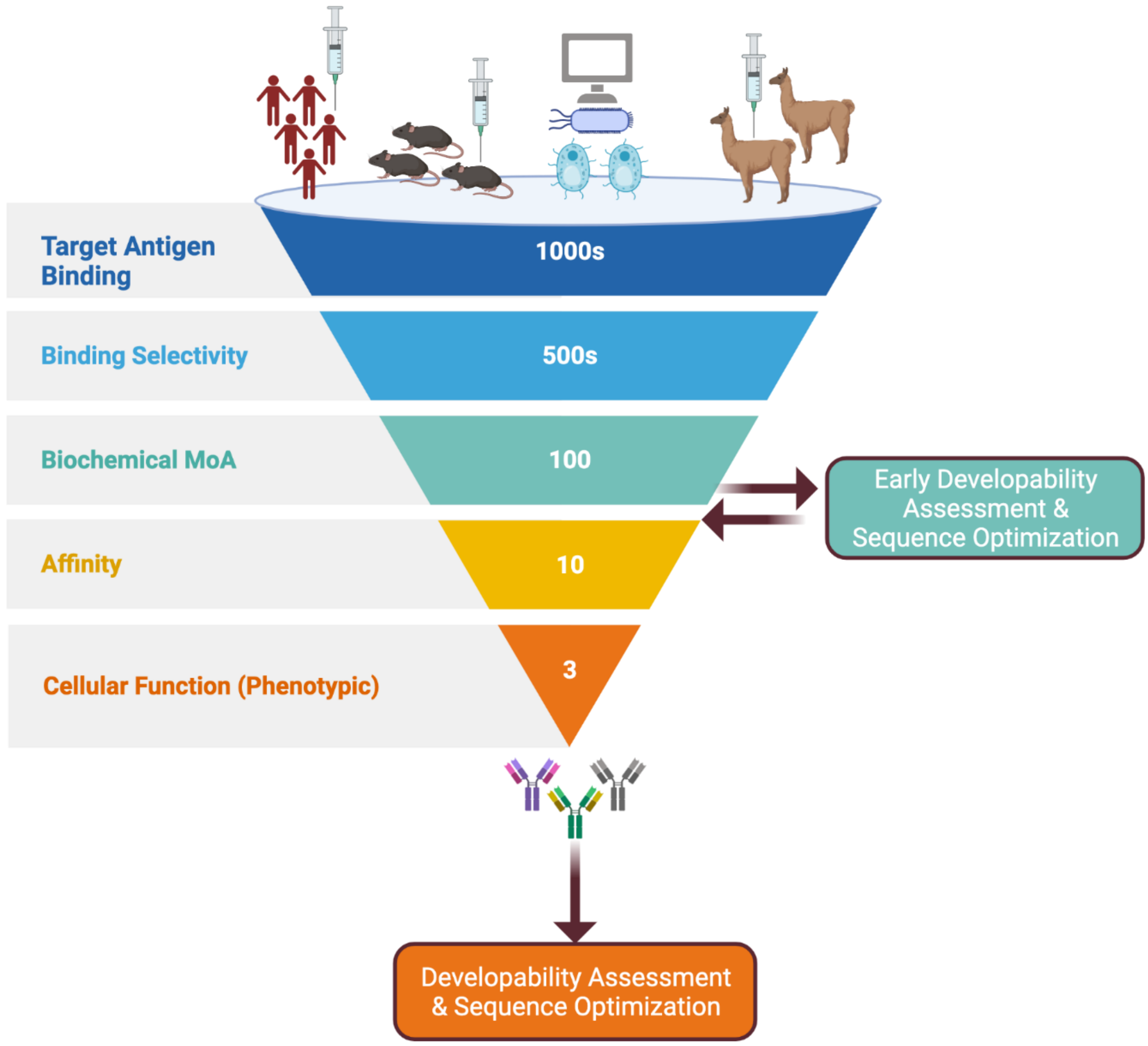

Figure 2. A generic hit discovery screening funnel. The initial diversity is rapidly restricted by successive high throughput assays, until the number of hits is sufficiently small that the most relevant, but typically cumbersome and low throughput, phenotypic assay can be performed to identify candidates prioritized for further, more expensive and challenging tests. Once hits with the desired activity are identified (orange), they are intensively optimized and engineered for improved potency and developability while also being scaled up and tested in *in vivo* disease models. An alternative approach is to perform sequence optimization earlier in the screening cascade, thus providing higher quality inputs to critical late screening assays (affinity and phenotypic).

A frequent outcome however, is that some of the lead candidates may be efficacious and others may be developable, but not all are *both* efficacious and developable. In this case, if the most advanced lead candidates fail to progress for any reason of efficacy or developability, the program may stall as the only alternatives are to return to the pool of unoptimized and partially characterized hits to search for a replacement lead molecule, or to re-engineer the sequence of the most advanced candidate for improved developability, followed by repeating all the downstream preclinical development steps. Either of these remedies entail a delay and an effort at which most organizations balk.

An alternative to the above unhappy scenario is to create a larger pool of fully or partially optimized hits which can serve to safeguard against attrition of the top lead candidates (Figure 2, “Early Developability Assessment & Sequence Optimization”). Furthermore, selecting the lead candidates for further characterization from a pool of optimized candidates allows for a fairer comparison between molecules and increases the probability that one or more candidates will meet all criteria for a lead molecule; this is because unoptimized hits may have poor properties (e.g. aggregation propensity or instability) detrimental to their performance in early assays in the screening cascade, obscuring a potential for greatness that becomes obvious only after optimization, as schematically represented in Figure 3.  We have even seen that sequence optimization based on developability criteria alone can improve affinity up to fifteen-fold (D. Nannemann, personal communication), thereby increasing the pool of hits that meet affinity criteria and can be assayed for the desired phenotypic effect.

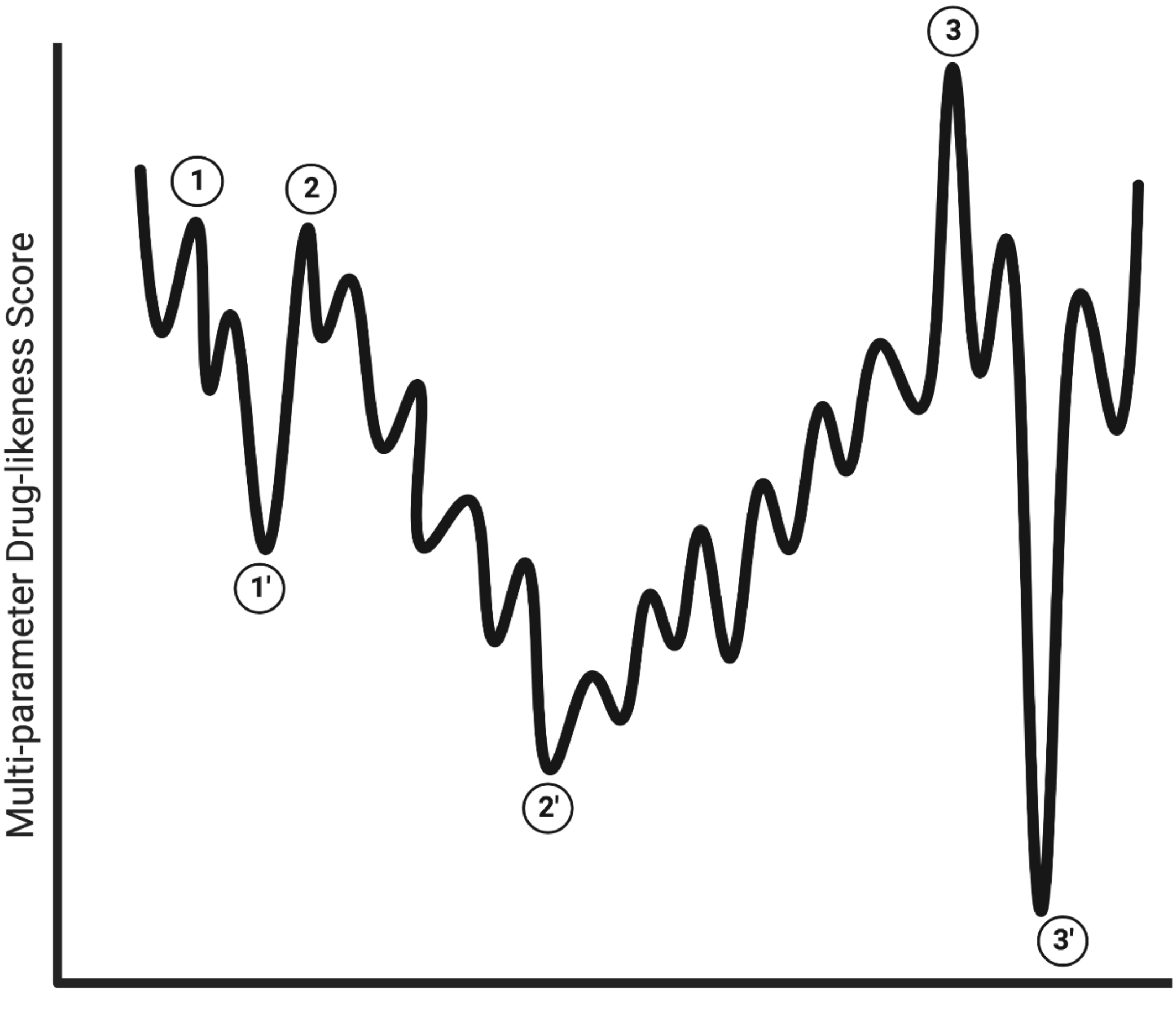

Figure 3. From confirmed hit to drug-like lead candidate. Three hypothetical hits from a screening cascade are symbolized as 1, 2, and 3. Their position on the y-axis represents how closely each conforms to a multi-parameter drug-likeness score that encompasses potency as well as developability and manufacturability aspects such as aggregation propensity, thermostability, hydrophobic or charged patches, viscosity, clearance, and immunogenicity. The sequence optimized version of each (1’, 2’ and 3’) will have a better drug-like score than its respective parent. Of note, the parental hits 1 & 2 which may score better than hit 3 (perhaps due to higher potency, for example, or fewer chemical instabilities, or less non-human sequence), may not be the ones that yield the most drug-like lead candidate. In certain cases, it may be impossible to fully optimize a certain hit (e.g. if potency and hydrophobicity are correlated). In other cases, a hit (3) might have more liabilities giving a poor score for drug-likeness, but if these can be easily rectified while maintaining potency, the "worst" parental hit may yield the "best" lead candidate (3').

However, many drug discovery organizations also balk at "investing" in optimization of hits that are not viewed as lead drug candidates (usually by a potency criterium), creating a sort of paradox. The best way to identify the most promising lead candidates is to compare diverse optimized hits, but optimization is only done for hits that already display the most lead-like attributes. The way out of this paradox is to change the organizational mindset regarding the value of early optimization. The mindset change is arrived at when the organization understands these two truths of drug discovery:

1. Early optimization is an investment that significantly reduces the probability of late-stage attrition of the eventual lead candidate, not a "waste" of effort on hits that will not become leads.
2. The investment in optimization does not have to be costly (in either time or money) if done judiciously and pragmatically

To convince the organization of truth #1, we suggest referring to concrete examples of later-stage failures as outlined in the introduction, in addition to internal examples from one's own organization; every drug discovery organization will have some confidential examples and it is wise to remember and learn from these. To avoid the fate of Cassandra of Troy [49], however, one must balance the lessons past failures by reassuring the organization that a path to success is achievable (truth #2). In a resource-limited setting, it is understandably tempting to adhere to a traditional screening funnel (Figure 2) which by its nature not only helps identify the few functional hits, but also limits the number of candidates that will undergo expensive scale-up and *in vivo* testing to very few. But it is due to the high cost to test the few candidates that it is imperative to ensure that the selected hits are the best possible lead candidates. Similar to the role of cheminformatics in small molecule hit triage [50], biopharmaceutical informatics can help reconcile the need to select the best hits with the reality of limited resources and time, by helping to select diverse and valid hits for optimization, and by reducing the resource required to fully optimize each selected hit. We outline here a resource efficient approach to optimization that can be applied to any drug discovery program in any setting, even when advanced computational tools such as those described in sections 9.3 and 9.4 are not available. This classical approach to optimization relies on access to any structural modeling software [51–55], general knowledge of antibody optimization, and a tight coupling of computation and experiment.

The first step in a pragmatic and efficient hit optimization procedure is to prospectively plan how many hits can realistically be sequence optimized. This will depend on the priority of the program in the portfolio, the throughput of each assay in the screening funnel, and the resources available to the program. Once the resource allocation is clear, the first critical scientific decision to be made is which hit sequences, of the (hopefully) hundreds that bind to the target, will be triaged and which will be discarded. Biopharmaceutical informatics is indispensable to making the best decision possible. After screening for desired selectivity and MoA, hierarchical clustering (generally on paratope sequence, but also on epitope binning if available) should be used to identify related clusters of hits. From these, one or several representatives from each cluster should be resynthesized and reconfirmed for target antigen binding. If the number of clusters exceeds the resource available for resynthesis and assays, sequence assessment should be used to identify and remove highly hypermutated hits, overly hydrophobic clusters, or clusters with inconsistent structure-activity relationships, as these are likely false positives that need not be tested further. At this point, if the number of clusters still exceeds the capacity for further testing, the number may be reduced by any other criteria such activity level, high throughput affinity assessment, protein engineering “intuition”, or random selection, until the number conforms to the previously agreed resource available. As long as the diversity is maintained, the program is safeguarded against attrition due to known and unknown causes as it progresses.

If sufficient diversity of biochemically active hits has been obtained in a given discovery project, the number of hits to be optimized will still likely exceed capacity if one attempts to do a comprehensive optimization designed to fully eliminate all risk from each sequence. Instead of such a “gold-plated” approach, we suggest a tiered approach (Figure 4). First, use structure-based drug design to optimize any liability where the sequence change will not affect potency. This typically entails germline humanization of residues that do not affect the conformation of the paratope [56] and will have the effect of stabilizing the antibody (which reduces the incidence of artefacts in screening assays) as well as reducing downstream immunogenicity risks. Second, include any optimization that will remove liabilities that are likely to interfere with the accuracy of read-outs from decision making low throughput assays (such as *in vivo* efficacy); beyond the humanization germlining that will be done anyway, this typically should include engineering any overt hydrophobic or charged patches, as these properties are prone to non-specific interactions and induce poor solution behaviors that interfere with assay read-outs and cause artefacts. Third, include any optimization that, should the hit develop into a contender for the program lead molecule, has a high likelihood of becoming a Critical Quality Attribute (CQA), and thus will almost certainly need to be removed from the sequence before manufacturing and clinical development. This will entail engineering the CDRs to remove chemically labile motifs and can be challenging to do in the absence of a co-crystal structure; it is also acceptable to postpone this optimization until a hit is prioritized as a lead candidate. Last, include optimizations that remove less likely theoretical risks or those that remove a risk with only moderate certainty of their beneficial effect. These include risk of immunogenicity which, despite the availability of excellent MHC-II binding predictors [57], still remains challenging to predict clinically; and risk of fast clearance or low bioavailability [25,26,47,58–60]. While it is preferable to de-risk as much as possible, it is not always pragmatic to de-risk extensively at the early hit optimization stage. One can safeguard against complete attrition due to less predictable factors by maintaining diversity; thus, allocating resource to partially optimizing more hits is preferable to fully optimizing fewer hits. This decision tree is stopped at the point where the number of sequence variants to be generated matches the capacity of the organization – the “good enough” optimization.

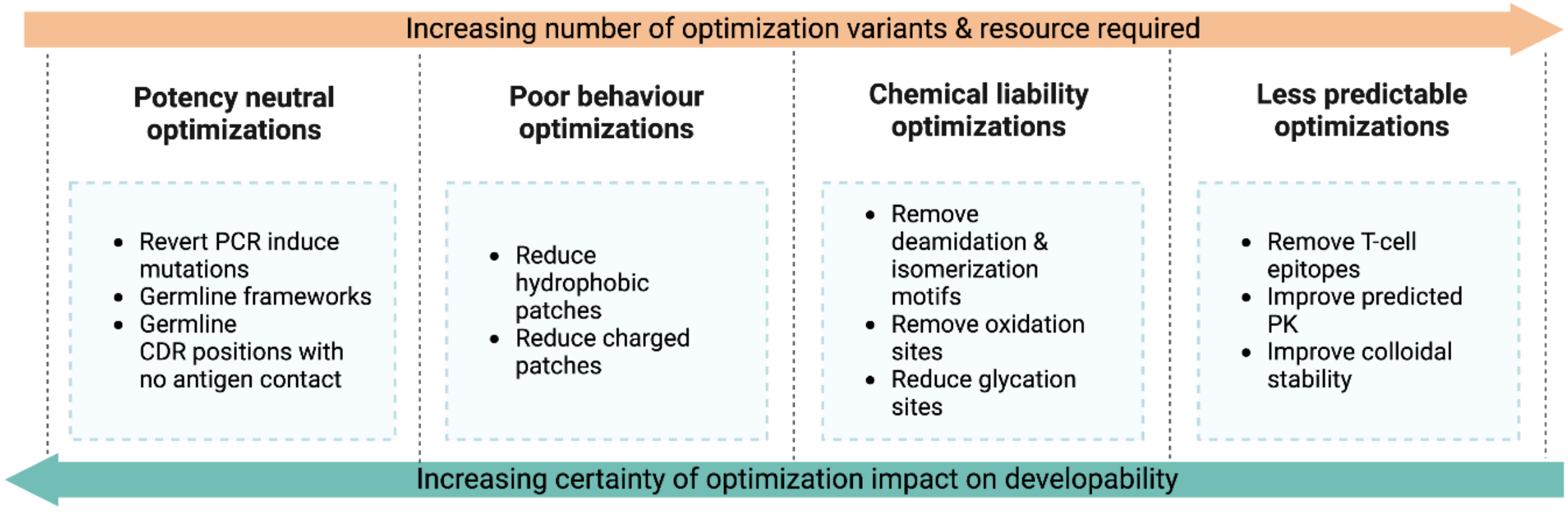

Figure 4. Tiered approach to prioritizing developability aspects to optimize. Towards the left are “must-have” optimizations that are relatively simple to execute and have a high probability of actually improving developability, as well as reducing false positive or false negative rates in pharmacological tests. The 3rd panel illustrates optimizations that positively impact the CQAs of a candidate; these are “must-have” but may be executed on fewer, prioritized hits. The rightmost panel includes “nice-to-have” optimizations.

It is also worth mentioning that investing in hit discovery platforms that produce fully human antibodies will obviate the need for humanization, and platforms based on specific immune repertoires frequently yield high affinity hits that obviate the need for extensive affinity maturation. Platforms such as these maximize the resource allocation for hit optimization as the optimization requirements are minimal and focused on manufacturability. However, if such platforms are not available, more extensive optimization can be done off the critical path, on fewer candidates, and one can focus on the most impactful optimizations for lead candidate selection in the first pass (Figure 4).

With agreement on the resources for hit optimization and a philosophy for a "good enough" optimization in place, all that remains is to actually optimize the selected diverse hits. Here again, biopharmaceutical informatics is critical in focusing limited resources on the most relevant and impactful optimizations that will support the best possible decision making for lead candidates; however, informatics alone cannot and should not guide the optimization. A strong collaboration with analytical sciences is key to a successful optimization in the classical approach. This "lean" approach is exemplified in Figure 5.

A.

| | VH | | | | |
|---|---|---|---|---|---|
| IMGT | 6 | 49 | 53 | 76 | 78 |
| Library | ab | cde | fg | hi | jk |
| # variants | 2 | 3 | 2 | 2 | 2 |
| VH-1.0 | a | c | f | h | j |
| VH-1.1 | a | d | f | h | j |
| VH-1.2 | a | e | f | h | j |
| VH-1.3 | a | c | g | i | k |
| VH-1.4 | a | d | g | i | k |
| VH-1.5 | a | e | g | i | k |
| VH-1.6 | b | c | f | h | j |
| VH-1.7 | b | d | f | h | j |
| VH-1.8 | b | e | f | h | j |
| VH-1.9 | b | c | g | i | k |
| VH-1.10 | b | d | g | i | k |
| VH-1.11 | b | e | g | i | k |

| | VL | |
|---|---|---|
| IMGT | 49 | 52 |
| Library | lm | no |
| # variants | 2 | 2 |
| VL-1.0 | l | n |
| VL-1.1 | l | o |
| VL-1.2 | m | n |
| VL-1.3 | m | o |

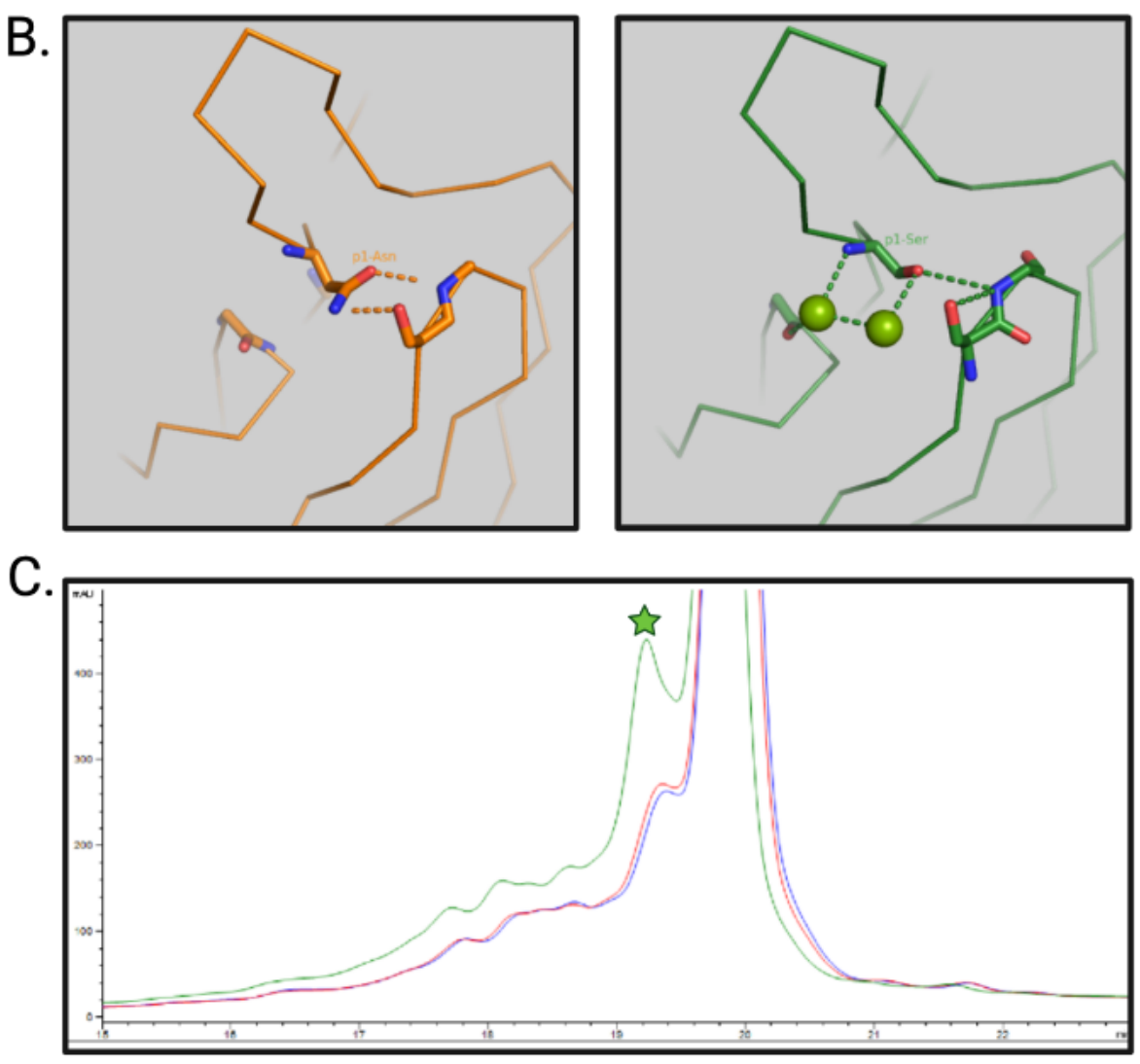

Figure 5A. Design of optimization variants. Upon structural and sequence analysis of a hit antibody, five VH residues and two VL residues are identified as sub-optimal, and possible substitutions are designed. The native and designed amino acid suggestions are schematically represented here as lower-case letters. The combinatorial table of designed and native variants is shown below. The variants designated as VH-1.0 and VL-1.0 represent the original parental sequence at each optimization position. The three positions shaded in green are covarying residues. For these residues, it is not necessary to sample all combinatorial possibilities. B. Structural analysis using a homology model of an asparagine residue implicated as a potential deamidation site (left panel) and substitution of the same position by a serine is shown. C. A chromatogram of a forced deamidation stress test is shown. The blue is the untreated control, red is the profile after three days at 37°C, and green is the profile after three days at 37°C at an elevated pH. The asterisk indicates the appearance of a well-known deamidation site located in the constant region.

When performing hit optimization, typically one or two mutations should be sampled at each sub-optimal sequence position. In order to arrive quickly at a sequence that is fully optimized at each position, it is preferable to express and test all combinatorial variants at once, however this may exceed the throughput of the screening assays if more residues are selected for optimization (right hand panels, Figure 4). An option to prevent a combinatorial explosion, is to test all heavy chain optimization variants only with the parental light chain, and vice versa; once identified, the optimal VH and VL can easily be combined in a transient transfection to yield the fully optimized hit. Additionally, one should use phylogenetic, structural and empirical analytical knowledge to further limit the number of VH or VL variants to be tested. For example, if residues are known to interact structurally and co-vary (green shaded residues in Figure 5A), rather than sampling all combinations it is advisable to restrict to only those combinations that are observed in nature and/or structurally compatible according to a homology model. When attempting to discover substitutions that eliminate the risk of a chemical liability such as asparagine deamidation, structural modeling can aid in reducing the number and type of substitution to be empirically assayed. For example, a homology model might reveal that a susceptible asparagine is involved in a structurally important hydrogen bonding network (Figure 5B, left panel). Substitution with a glutamine, while conservative from a physicochemical perspective, would disrupt this network whereas substitution with a serine can preserve the network – although it is necessary to model waters in order to predict this (Figure 5B, right panel), highlighting the importance of expert computational structural review of the homology models and proposed substitutions. Finally, experimental analytics can also be used to prioritize which substitutions to add to the combinatorial set of variants. Figure 5C demonstrates that in a three-day forced degradation test on an asparagine containing antibody, no deamidation in the V-region is seen. This does not mean that the residue in question should not be optimized eventually, as it may show chemical instabilities when tested in a more stringent assay; however, it does suggest that resource allocation to optimizing this position should be postponed until downstream tests reveal that residue is truly labile. Additionally, if this hit is not selected as the lead candidate, it will not be necessary to optimize this position at all; the three-day test already confirms that the antibody will be stable enough to give reliable decision-making data in screening and confirmatory assays. Finally, although it is important to reduce the number of optimization variants to be tested to a lower number suitable to the early phase of the program, it is wise to include at least an alanine mutation at any potentially labile positions identified in the CDRs, as these positions have a higher risk to become CQAs if the selected hit becomes the lead candidate. While alanine substitutions are unlikely to become part of the final optimized lead sequence, they do provide a rapid and reliable way of assessing which CDR residues are directly involved in antigen binding – information which is highly valuable in the absence of a co-crystal structure and will enable a rapid final optimization should the hit become the lead candidate.

This classical approach to developability optimization has the advantage that it is available to any drug discovery organization with a screening team, an analytics expert and an expert protein modeler. However, in the absence of structural modeling expertise, when under time constraints due to competition, or when the initial hits require more extensive optimization, this approach becomes limiting. Therefore, we describe better contemporary and emerging approaches below which hold the promise of enabling more hits to be extensively de-risked, with less resource.

### 9.3 The contemporary approach – engineered libraries towards improved developability

An alternative for the design of a few specific sequence optimization variants of antigen-specific antibodies obtained from repertoires of naïve, immunized or synthetic subjects is the generation of display libraries that are by design engineered to produce hits with improved developability, ideally in combination with NGS and (optionally) deep learning approaches. Historically, antigen-specific sequences from display libraries are identified using fluorescence activated cell sorting (FACS "panning"), followed by Sanger sequencing of a few selected clones with increased binding signals. Meanwhile, NGS or "deep" sequencing technologies have been established that reveal how entire sequence spaces and variations of antibody libraries recognize antigens.[61–63]. Typically, antigen-binding cells are isolated through multiple rounds of display panning against the target antigen and the binding/non-binding populations are subjected to next-generation sequencing. As an advantage, NGS analysis of sequence pools before and after panning can offer new ways of finding rare but highly enriched potent binders and sequence motifs that might not be identified by random picking and Sanger sequencing.[64] The sequence and sequence-activity-relationship (SAR) spaces obtained from NGS can be used to train AI/ML models to predict new sequences with even further improved binding affinities or developability properties (Figure 6).

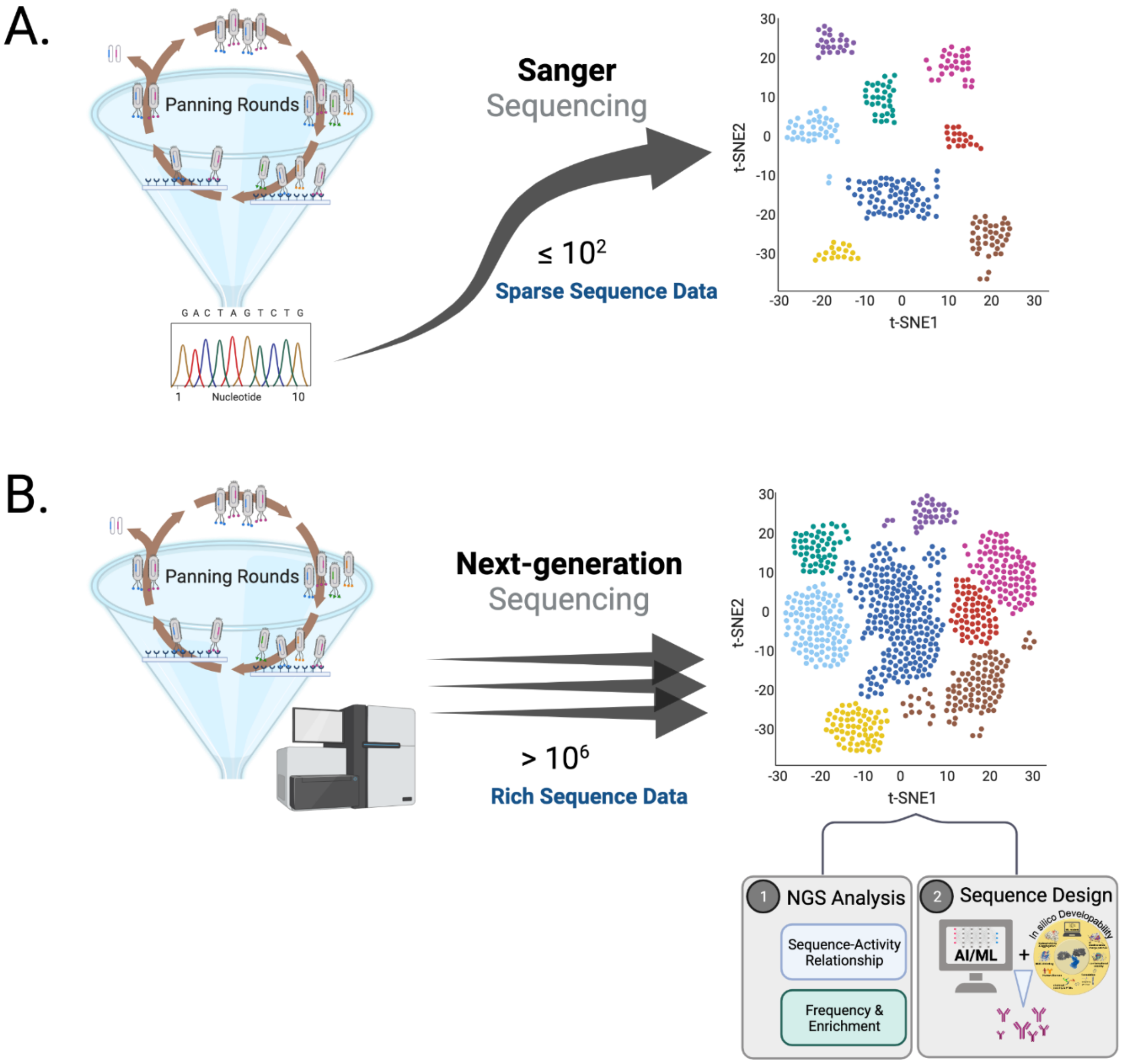

Figure 6A. Process of hit identification from display libraries based on multiple rounds of bio-panning and random clone selection for Sanger sequencing. Typically, only a low number of clones can be sequenced, providing the risk that potent sequences with (more) promising sequences or even sequence clusters with favorable developability properties will be missed. B. NGS analysis of sequence pools before and after panning allows assessment of entire sequence spaces of potent binders. Using SAR, frequency and enrichment analyses in combination with *in silico* developability predictions, it is possible to rationally identify potent binders with a favorable developability profile. In addition, the sequence pools can be subjected to AI/ML approaches to predict new sequences within even further improved potency and developability properties.

### 9.3.1 Design of specific hit optimization display libraries in combination with AI/ML approaches

Instead of designing specific sequence optimization variants as described in section 9.2, combinatorial hit optimization display libraries can be produced where sequence diversifications are typically introduced in CDR regions [65].

As an example, Liu et al. [66] generated a synthetic library containing ~$10^{10}$ ranibizumab variants that were randomized in CDR-H3 for three rounds of phage display panning. Based on the frequency distribution and enrichment analysis obtained from NGS analyses of different panning rounds, ML models were trained with a two-stage approach, by first modeling antibody affinity and specificity in terms of sequence enrichment with an ensemble of neural networks and then optimizing it with gradient-based optimization. These models were able to predict new CDR-H3 sequences of ranibizumab that were superior to the sequences in the training data set. In another study, Saka et al. [67] employed deep generative models based on NGS derived sequences from different panning rounds of a combinatorial library diversified in CDR-H1, -H2 and -H3 and FR1 of a kynurenine binding antibody. The affinities of newly designed sequences were over 1800-fold higher than for the parental clone.

In a further study, Mason et al. [68] performed single-site deep mutational scanning within CDR-H3 of trastuzumab and subjected the antigen-binding vs non-binding samples to NGS to calculate enrichment scores and guide the rational design of a combinatorial mutagenesis library of $5 \times 10^4$ variants. This library was screened for specificity to the antigen HER2, again followed by next-generation sequencing. All binding and non-binding variants were used to train deep neural networks that were then applied to design and rank a set of $10^8$ virtual sequences. To explicitly account for the aspect of developability, the set of predicted binders were then subjected to sequence-based *in silico* filtering steps to optimize for developability parameters such as viscosity, clearance, solubility and immunogenicity, resulting in nearly 8,000 antibody sequence variants predicted to have more optimal properties than the starting trastuzumab sequence. Further biophysical characterization of top sequences revealed antibodies with comparable or better properties than trastuzumab for expression and thermal stability, and one candidate that was substantially de-risked for immunogenicity.

Whereas Mason et al. applied *in silico* developability scoring to identify binders with favorable developability properties, Makowski et al. very recently presented an approach where a CDR combinatorial mutagenesis library of ~$10^7$ emibetuzumab variants was sorted for high and low levels of affinity and non-specific binding to two polyspecificity reagents [69]. Different ML models were then trained on the NGS-enriched libraries for both affinity and (non-)specific binding. In agreement with previous work, the authors demonstrated a high correlation between high affinity and non-specific binding of emibetuzumab variants. The model predictions were used to identify those antibody mutants in the library that maximize antibody affinity to different extents while minimizing tradeoffs due to reduced non-specific binding. Additional models, that used Unified Representation (UniRep) or physicochemical features as antibody descriptors, were built and then applied to design new variants into a novel mutational space that was not covered by the mutations of the training data set. Production and experimental profiling indeed revealed new sequences with even greater improvements in affinity and specificity than was possible in the experimentally sorted libraries.

The number of reports of such AI/ML approaches in combination with NGS and biopanning is rapidly increasing [70,71], demonstrating their general potential to transform the Biologics drug discovery process. In conclusion, these studies demonstrate how AI/ML models can be built based on sequence enrichments that are observed in antibody libraries through different rounds of biopanning. As demonstrated by Makowski et al., such models can be trained and used for the co-optimization of multiple optimization parameters, such as antigen binding and non-specificity. Of course, these concepts can also be further generalized towards additional optimization parameters, for example to optimize antibodies for cross reactivity against the antigen of relevant pharmacological or toxicological species. Further developability parameters can be considered already in the design of the combinatorial sequence space, for example by excluding sequence combinations that are predicted to show unfavorable predicted physicochemical (e.g. liabilities, PTMs, hydrophobicity, viscosity, charged patches, etc.) or immunogenic (low human-likeness, MHC-II binding) properties. Alternatively, or in addition, such *in silico* predictions can be used to post-filter the ML-predicted variants to select sequences with the best overall affinity and developability profile, as demonstrated by Mason *et al*.

### 9.3.2 Design of diverse hit identification libraries

Different antibody library design approaches towards improved developability have been excellently summarized in the last edition of "Current advances in biopharmaceutical informatics" [33] and in publications cited therein. Here, we briefly summarize the different strategies for the design of diverse libraries for hit identification, and in addition highlight some very new advances in this field.

**Immune and naive libraries**. Libraries derived from antigen exposed or naive humans or humanized animals are based entirely on naturally occurring sequence diversity and are generally considered to bias obtained sequence spaces towards high expression and low immunogenicity. However, there is no need or driving force for the immune system to construct antibody sequences towards developability properties that are relevant for the pharmaceutical industry, such as the long-term physical and chemical stability. Therefore, antibody libraries obtained from naive libraries often need further optimization towards required biophysical and chemical stability properties.

**Synthetic and semi-synthetic libraries**. Synthetic or semi-synthetic libraries combine natural diversity for certain aspects of the library with *in silico* design, often with regard to further improved developability properties [72], such as high thermal stability, reduced aggregation propensity, low occurrence of PTM or chemical liability sites. However, synthetic diversity may create artificial complementarity-determining region (CDR) sequences that fold poorly, since the design strategy often includes the combinatorial enumeration of amino acid mutations in specific positions of otherwise fixed antibody scaffolds based on positional frequency analysis (PFA). This procedure, however, ignores how residue types interact to form stabilizing interactions such as hydrophobic, hydrogen or ionic bonds, potentially resulting in many library members that fold and express poorly. One recent approach for a rational design of a semi-synthetic library that explicitly considers developability and functional compatibility of antibody framework and CDR regions was described by Teixeira et al. [73]. In their library design, HCDR3s were amplified directly from B cells from 10 healthy adult human donors and grafted onto four paired human frameworks. These were derived from a diverse panel of well-behaved (“developable”) clinical antibodies, based on known biophysical characteristics [23] and at the same time cover different germline families and thereby assure structural and sequence diversity in the library to improve the ability to select binders against different antigens. Finally, to optimize for developable sequences encoding CDRs able to express well within the chosen scaffolds, natural human CDR-L1-3s and CDR-H1-2s as found in a human NGS dataset were first purged of defined sequence motifs related to chemical instability, PTMs, polyreactivity and surface hydrophobic/aromatic patches. In contrast to using degenerate oligonucleotides, these sequences were produced using oligonucleotide array-based synthesis, thereby reducing the combinatorial library space and increasing the likelihood of proper folding due to an inherent compatibility of the CDRs within the variable light or heavy chain. In the next step, these CDRs were filtered by yeast display as single CDR libraries as further criterion to eliminate sequences negatively impacting expression, folding and display. Finally, all CDRs were combinatorially assembled into a single-chain variable (scFv) format. The general performance of this library approach was successfully demonstrated by the discovery of a large number of unique, highly developable antibodies against four clinically relevant targets with affinities in the sub-nanomolar to low nanomolar range.

Very recently, another novel library approach for the high-throughput *de novo* identification of humanized VHHs following camelid immunization was implemented [74]. For this, VHH-derived CDR3 regions obtained from a llama (*Lama* glama), immunized against a specific target, are grafted onto a humanized VHH backbone library comprising sequence-diversified CDR1 and CDR2 regions similar to natural immunized and naïve antibody repertoires. Importantly, these CDRs were tailored towards favorable *in silico* developability properties, by considering human-likeness and excluding potential sequence liabilities and predicted immunogenic motifs. Target-specific humanized VHHs against this specific target were readily obtained by yeast surface display. By exploiting this approach, high affinity VHHs with an optimized potency and developability profile can be generated that do not require any further sequence optimization. In a further study, the screening pools obtained from different panning rounds against were submitted to NGS derived enrichment analyses (manuscript in preparation). For four highly enriched clusters, deep generative models were trained based on the sequences obtained after panning and used for the sampling of new sequences. Top-ranked sequences were subjected to sequence- and structure-based *in silico* developability assessment to select a set of <10 sequences per cluster for synthesis. As demonstrated by binding measurements and profiling in screening assays for early developability assessment, this procedure might represent a general roadmap for the fast and efficient discovery and design of potent and readily optimized VHH hits with favorable early developability properties directly from screens of immunized lama repertoires.

In future applications, such library spaces might be even further tailored using additional or alternative *in silico* developability descriptors, such as the computed isoelectric point (pI) [48], or other scores that are believed to be predictive for relevant developability properties as described in the next section.

**Libraries derived from deep generative models**. In recent years, several publications have described the application of deep generative models, a combination of generative models and deep neural networks, for antibody or single variable domain on a heavy chain (VHH) antibody library design [75–78]. A generative model analyses the distribution of the training data itself, and provides an estimate how likely a given example (e.g. antibody sequence) is. Deep generative models are neural networks trained to approximate complicated, high-dimensional probability distributions and can be used to create new samples (sequences) from the underlying distribution. Translated to the task of antibody library design, deep learning offers a route to capture the complex relationships between protein sequence and structure behavior and thereby opens the door to design sequences that are stable and functional. As a concrete example, such models learn from an underlying dataset (for example NGS datasets from human donors), which antibody framework and CDR sequences and modifications are (structurally) compatible with each other and will ideally only suggest new sequences that will results in functional antibodies, which will express properly. Since the training data can be derived from huge datasets of human antibody sequences, the deep learning method implicitly learns the rules for constructing (only) human-like sequences, and the resulting library will ideally retain typical human repertoire characteristics with a low risk of immunogenicity. However, not all predicted sequences will necessarily show favorable properties with regard to further developability parameters, since - as mentioned above – the human immune system does not optimize sequences for all developability properties that are relevant for a pharmaceutical drug product. Thus, the resulting sequences may still contain chemical liability or PTM motifs, or might show non-favorable solution behavior at high concentrations. Options towards the design of more developable libraries might be (i) the "pre-cleaning" of the sequences from the "training" data set based on suited *in silico* developability filters or (ii) the post-filtering of the designed library using the same *in silico* scores. However, such an approach might not be applicable for huge libraries, in particular in cases where the *in silico* property calculation includes computationally expensive steps, such as 3D-model generation or even conformational sampling. In such cases, it might be feasible to add a further step of so-called "transfer learning" after an initial (purely sequence-based) deep generative learning step. Transfer learning is a continued training of a model with a small subset of sequences with specific desirable characteristics (as for example identified by proper *in silico* filtering using computationally "expensive" methods) and might thereby be suited to bias the properties of the generated antibody sequences toward improved developability properties relevant for the pharmaceutical industry (e.g. physical & chemical stability and immunogenicity).

Such a deep learning approach, including transfer learning, was described by Amimeur et al [75]. An initial library that was trained using a Generative Adversarial Network (GAN) to learn the rules of human antibody formation on a set of over 400,000 human antibody sequences. Through transfer learning on small library subsets with desired *in silico* properties, the finally designed library could be biased to generate molecules with key properties of interest such as a defined (*in silico*) pI range, CDR-H3 length, computed negative patch area and lower predicted MHC class II binding.

Whereas these libraries were designed as generic hit identification libraries, applicable for a diverse set of targets, another study reported by Lim et al. [79] demonstrated that deep learning models can also predict new antibody binders for specific targets when trained on diverse sequences binding against these targets. Sequences that had been obtained from yeast display and NGS analysis against PD-1 and CTLA-4 were classified into binders and nonbinders based on sequence reads and enrichment factors from FACS sorting. The antibody CDR-L3 and CDR-H3 sequences were encoded into antibody images representing their BLOSUM substitution scores, which were then used to build and train convolutional neural network models for specific (V gene) sequence clusters to classify binders and nonbinders. The *in silico* generated sequences were similar to the sequences of the training data set, but interestingly also included mutations that were not present in these training sequences. The machine learning procedure reported in this study was designed to identify potential new binders against the investigated antigens. As perspective, future studies might combine this approach for library design with the different above-described strategies to tailor such libraries for the prediction of antigen-specific binders with an optimized *in silico* developability profile.

To incorporate these contemporary approaches as general and essential part of hit discovery towards developable hits, organizations should ensure continuous and close collaboration of protein engineers, deep sequencing experts, data scientists and machine learning experts.

### 9.4 The emerging approach - *De novo* design of developable antibody therapeutics

As exciting as the contemporary approach that focuses on engineered libraries with improved developability is, it relies on target-specific datasets (either bespoke or from the public domain), which are then used to train models to optimize libraries. With recent advances in deep learning technology, neural networks are being applied to tasks such as sequence design, fold recognition, paratope prediction [80], epitope prediction [28,81] and structure prediction [51,82,83]. However, the advent of more advanced AI-based approaches for generating new antibodies from scratch, i.e. without relying on the natural immune system or using a template from existing antibodies, has shown immense potential in designing sequences for target antigen binders. These binders are aimed to have properties like high affinity, specificity, stability, and excellent developability profile. This *de novo* antibody design approach bypasses all the time-consuming data generation steps for model training for each discovery campaign for novel and optimized antibody-based therapeutics. Lately, we have seen generative models [84,85], that approximate the distributions of the data they are trained on. These have garnered interest as a data-driven way to create novel proteins. Unfortunately, most protein-generators create 1D amino acid sequences, making them unsuitable for problems that require structure-based solutions, such as designing protein-protein interfaces like the antigen-antibody complex. In the last few years, two methods shined their capabilities where with known target antigen structure and epitope residues, they either used AI-generated immunoglobulin-like backbones [84] (figure 7 pathway 1a) or used deep learning to virtually screen the whole structural antibody space for binders [86] (figure 7 pathway 1b).

The first method listed above utilizes Variational Auto-Encoders (VAEs) with a unique loss function for xyz-coordinate generation. This serves as the solution for distance matrix reconstruction and torsion angle inference while preserving the desired invariances. In other words, by constraining and optimizing a protein structure in the VAE latent space, it becomes feasible to specify any desired structural features while the model creates the rest of the molecule. As an example, pathway 1a involves Ig-VAE that can perform constrained loop generation, towards epitope-specific antibody design [84]. After the backbone construction of this model via constrained optimization in the latent space of a generative model, it becomes a protein sequence design problem where given a protein backbone structure of interest, an amino acid sequence that will fold to this structure.

Historically, physics-based methods such as Rosetta treat sequence design as an energy optimization problem, searching for the combination of amino acid identities and conformations that has the lowest energy for a given input structure [54]. In past few years, the evolutionary relationship has provided an alternative where the constraints on protein structure are derived from bioinformatics analysis of the evolutionary history of proteins, homology to known structures, and pairwise evolutionary correlations [54,87]. As we know, the protein sequence space is large, discrete, and sparsely functional, where only a small fraction of sequences may fold into stable structural conformations. Adding on to the complexity, immunoglobulins are a class of proteins that follow different evolutionary pressures as compared to other structures in the Protein Data Bank (PDB) and thus re-training on the multiple sequence alignment (MSA) for immunoglobulins is required to predict the structures of CDR loops [78]. Furthermore, the intrinsic flexibility in CDR loops is another reason for this class of proteins to be challenging for automated and efficient exploration of *de novo* design space.

However, recently, autoregressive and non-autoregressive generative model-based approaches have been gaining the limelight. One such example is "deep manifold sampling," which can be used for accelerating function-guided protein design [88]. By combining a sequence denoising autoencoder (DAE) with a function classifier trained on 0.5M sequences with known function annotations from the Swiss-Prot database, such a sampler can generate diverse sequences of variable length with desired functions. Parallel to this non-autoregressive method, an autoregressive language model ProteinMPNN [89,90], can also be used to generate all possible candidate amino acid sequences given monomeric protein backbones without the need for compute-intensive explicit consideration of side chain rotameric states (figure 7 pathway 1a). These sequences can be folded into respective structural folds using advanced methods like AlphaFold2 [87] and ESMFold [91]. Even though these methods are developed for all proteins, these have potential to be applied to immunoglobulins with ease.

An alternative to above-mentioned multi-step protocol (pathway 1a) a more concise method, as shown in pathway 1b (figure 7), uses a virtual screening approach to obtain the structure of antibody binders for the target in question. This is done by ranking structural antibody libraries by their predicted binding affinities against the pre-defined target epitope, thereby allowing direct antibody discovery against an epitope region of interest [86]. With great capabilities comes great responsibilities. The success of such virtual screening campaigns greatly depends on the prediction accuracy of the docking and scoring algorithms for antibody ranking. As recently reviewed [92] and evaluated [93], structural prediction of protein-protein complexes using docking and AI-based approaches still leaves room for improvement, especially in antibody-antigen complexes derived from structural antibody homology-based models. Similar to any docking problem, aspects like the accurate prediction of flexibility, the role of water and further factors might represent potential challenges [94] in this case as well. Therefore, first success stories about prospective hit identification of new antibodies from AI-based docking approaches of huge libraries are eagerly awaited.

Encouragingly, the first application of deep learning for structure-based virtual screening of antibodies (DLAB) has recently been described in a retrospective study by Schneider *et al* [86]. DLAB utilizes a convolutional neural network trained on rigid-body docking poses of modeled antibody structures in complex with antigen epitopes to improve the ranking of docking poses from the ZDock algorithm [95] and, in combination with docking scores generated by ZDock, for the prediction of antibody–antigen binding. On datasets of known antibody-antigen pairs, the dataset of known binders could be enriched among the top-ranked scorers, with a significantly higher success rate when docking crystal structure conformations compared to homology models.

Once the sequence and structures of the candidate binders are predicted, their developability needs to be evaluated on several factors like binding affinity, human-likeness, chemical liabilities, PTMs, MHC-II binding, aggregation propensity, thermostability, PK, etc. using suitable *in silico* developability assessment [29–34,36–39]. This focused set of binders can then progress to an antigen binding screen more confidently and later be developed into an antibody format of interest. The overarching goal for this field is to incorporate all aspects of developability as a learned index at training stage of the sequence design stage (Stage 2, figure 7) so that all generated sequences can bypass the need for post-hoc *in silico* filtering for any liabilities.

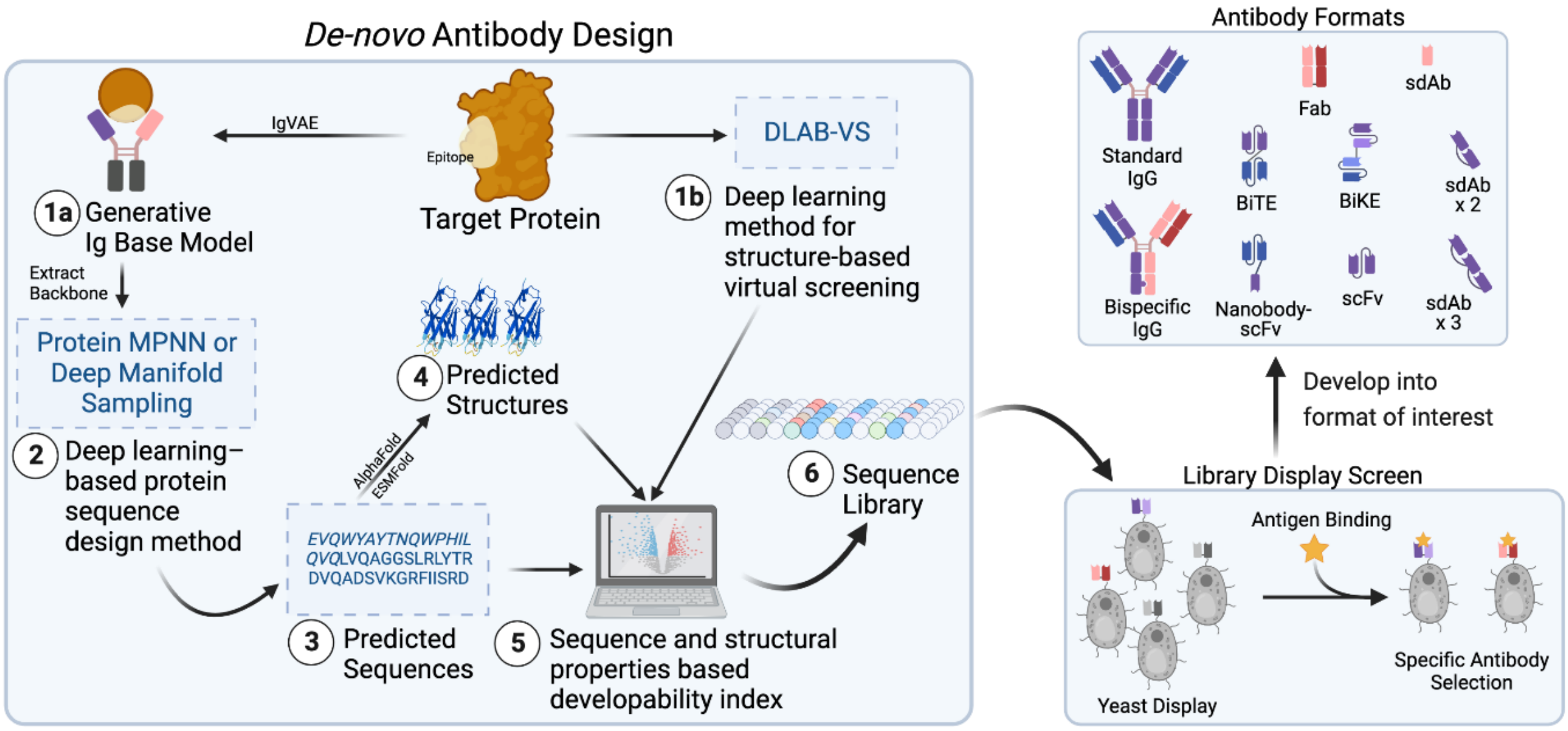

Figure 7. *De-novo* antibody design for known target and epitope. Pathway 1a involves a multistep procedure starting with developing a base structural model of the binder that is complementary to the epitope and is thus expected to bind with high affinity. Backbone of this model is then used by Protein MPNN to generate a set of diverse sequences that can fold into the same backbone while maintaining the affinity for the target. Alphafold2 can be used fold these predicted sequences. ESMFold can align the predicted structure for the designed sequence with the original backbone. Pathway 1b, on the other hand, uses DLAB, a deep learning based virtual

screening protocol for finding binders based on predicted binding affinity for the target in question. These pathways merge at step 5 where they are evaluated for their developability profile before progressing to antigen binding-based selection. These binders can then be molded into antibody format of interest.

## 9.5 Conclusions and Outlook

Historically, antibody discovery and optimization were often pursued following a reactive rather than a proactive approach: Antibody sequences obtained from traditional screening approaches were mainly optimized on binding affinity, human-likeness and chemical liabilities, followed by developability assessment of the best hits. Since these assessments were typically performed after sequence optimization, they served only to identify and flab sub-optimal developability properties in lead molecules. The issues resulting from these sub-optimal properties were thrown over the proverbial fence to downstream functions (DMPK, non-clinical safety and CMC). These functions then attempted to compensate for sub-optimal candidate properties through optimization of the downstream process development and dosing regimens, imposing delays in development, increased costs and finally a huge risk for the project to achieve approval for First in Human and further clinical studies. Thus, in recent years the pharmaceutical industry has come to the realization that the optimization of antibody-based biologics should be treated as a holistic multi-parameter challenge that considers multiple developability aspects (beyond potency and specificity) as defined by the specific target product profile of a project as early design parameters. Prioritizing early optimization for developability or applying emerging options to generate drug candidates de-risked for developability by design, as described in this chapter, should be viewed as investments that will pay dividends in reduced downstream costs and attrition. The contemporary and emerging artificial intelligence and machine learning (AI/ML) approaches are highly auspicious to transform biologics discovery, using less resource for optimization and promising to deliver truly de-risked candidates with lower attrition rates directly from hit discovery or by *de novo* design. We have to be aware, however, that the general success and impact will critically depend on the general predictivity of the *in silico* predictors. While great progress has been achieved in implementing novel AI algorithms, there remains a gap in the availability of specific relevant experimental data that would allow accurate prediction of clinical success of antibodies. Whereas there are, for example, large amounts of data from *in vitro* and *in vivo* assays to assess PK properties in preclinical species, the available landscape of antibodies with curated human PK data is currently by far too low to establish and validate robust and truly predictive models for human PK [47]; additionally the clinical predictivity of *in vitro* assays or even human FcRn transgenic rodent *in vivo* PK assays is not as reliable as the predictivity enjoyed by the equivalent small molecule assays [96]. Furthermore, even the available preclinical data is largely siloed within the “dark matter” of proprietary pharmaceutical company data. Similarly, the amount of experimental viscosity data available to individual pharmaceutical companies from literature or inhouse experiments is rather in the range of tens to hundreds instead of thousands or more. Furthermore, viscosity behavior may critically depend on specific assay settings or formulation conditions, thereby increasing the risk that *in silico* models trained on these data are outside the applicability domain of new sequences in specific formulations. Nevertheless, although several “late-stage” parameters cannot yet be accurately predicted, early *in silico* assessment can have a valuable impact in providing an educated rationale for sequence ranking and filtering, triggering of relevant and predictive assays, and design of potential backup sequences as early as possible. Such cycles of property prediction and experimental testing will ideally increase the amount of available data and continuously improve model robustness and accuracy. Concomitantly with this continuous improvement of model accuracy, it is essential to ensure each prediction is accompanied by a relevant confidence metric, to guide appropriate use of models. Finally, whereas we have already seen significant progress in property prediction for monospecific antibodies (mostly IgG1s), this progress urgently needs to be extended to *in silico* approaches for developability prediction of different antibody classes (e.g. IgG2, IgG4 and IgM), different multi-specific formats, complex fusion proteins, and antibody-drug conjugates as next-generation antibody-based drugs. While the developability data on different biologicals are currently mainly shared through the peer-reviewed literature, we believe that consortia, public-private partnerships, and the establishment of a comprehensive biologics data commons are essential to the future. Through the increasingly open sharing of data the entire industry, as well as patients and payers, will benefit from improved models of developability and the concomitant lower attrition rates.

**Acknowledgements**

We thank Jessica Dawson and Navneet Singh for providing representative deamidation data (Figure 5C); Lukas Friedrich for valuable discussion on deep learning approaches; Joleen White, David Nannemann and Sandeep Kumar for suggestions on case studies to illustrate developability challenges and solutions.